# Structure-Composition Correspondence in Crystalline Metamaterials for Acoustic Valley-Hall Effect and Unidirectional Sound Guiding

SIMON YVES[1], GEOFFROY LEROSEY[2] AND FABRICE LEMOULT[1,(a)]

[1] *ESPCI Paris, PSL University, CNRS, Institut Langevin - 1 rue Jussieu, 75005 Paris, France*
[2] *Greenerwave, ESPCI Paris Incubator PC'up, 6 rue Jean Calvin, 75005 Paris, France*

**Abstract** – Recently, the domain of topological insulators in condensed matter physics has been teeming with intriguing and very exciting discoveries. Notably, the capacity of guiding currents towards specific directions according to the spin of the travelling electrons has a great potential for electronic devices. This new phenomenon has been transposed to the classical domain in electromagnetics and acoustics, unveiling the pseudo-spin locking of guided waves. However, these macroscopic analogues are photonic/phononic crystals which are intrinsically wavelength scaled. In this work, we realize a genuine acoustic analogue of the valley-Hall effect in the audible regime using a lattice of soda cans. The crystalline description of this very simple metamaterial allows us to demonstrate experimentally the unidirectional excitation of sound guided at a scale much smaller than the wavelength of operation. These results not only open the tantalizing valley-topological phenomena to the audible regime but also allow to envision compact applications for acoustic manipulation.

**Introduction.** – Sound manipulation is of great interest in numerous domains of application such as acoustic isolation, ultrasonic imaging, sonars and communication for example. In the last decades, it has been demonstrated that phononic crystals [1, 2] permit to tailor the propagation of sonic waves with great precision. However, their wavelength-scaled structure makes them bulky devices, notably for low frequencies. Metamaterials, on the other hand, are man-made composite media which are structured on a scale much smaller than the operating wavelength [1, 3, 4, 5]. In the case of a resonant unit cell, these systems are named locally resonant metamaterials [6, 7, 8] and are usually described in terms of effective parameters which neglect or at least mitigates their sub-wavelength structural peculiarities. These homogenization procedures have led to the discovery of intriguing new phenomena such as negative index and acoustic cloaking [9, 10, 11, 12, 13], near-zero density media [14] and deep subwavelength imaging [15].

Nevertheless, due to the resonant nature of its constituents, multiple scattering occurs within the unit cell of such locally resonant metamaterials [16, 17]. It means that both their structure and their composition, albeit at a deep subwavelength scale, play a major role on the sound propagation. Hence, adopting such a microscopic approach has led to the obtention of negative refraction at a macroscopic scale [16] as well as topological properties [18], opening the era of so-called crystalline metamaterials. Moreover, the waves propagating in these systems follow the physics of the polariton [17, 19, 20] meaning that the sound is confined in the close vicinity of the resonators. This allows to obtain field maps experimentally and makes these media great table-top platforms for the study of fascinating solid-state physics phenomena at a macroscopic scale.

In this letter, we characterize numerically and experimentally an acoustic analogue of the quantum valley-Hall effect using a very genuine metamaterial made of soda cans, in the audible regime. We first establish a clear connection between structure and composition comparing the honeycomb and Kagome lattices. From there, inspired by a previous work in the microwave domain [21], we use this correspondence to straightforwardly induce a circular crystalline polarization, so-called valley degree of freedom, by modifying slightly the subwavelength structure. In that way, it differs from the work of [22] which exploit the composition of the unit cell. We then use it to make specific interfaces which present guided modes carrying also a crystalline polarization linked to their direction of propagation. Last, but not least, this intriguing property is highlighted experimentally by the straightforward rotational excitation of such directional subwavelength acoustics waveguides.

**Correspondence between Honeycomb and Kagome lattices of soda cans.** – Several macroscopic analogues of the quantum valley-Hall effect have been demonstrated for acoustic waves, with phononic crystals [23, 24, 25, 26, 27, 28, 29, 30, 31, 32], but also for elastic waves in structured plates [33, 34, 35, 36, 37] and even at the scale of a chip [38]. Their creation starts by the presence of Dirac cones within the band diagram of the sample, and then by breaking them appropriately. In this study, we work in the airborne acoustic domain with a very simple locally resonant metamaterial whose building block is a mere soda can. It is a low-loss Helmholtz

[(a)]E-mail: `fabrice.lemoult@espci.psl.eu`

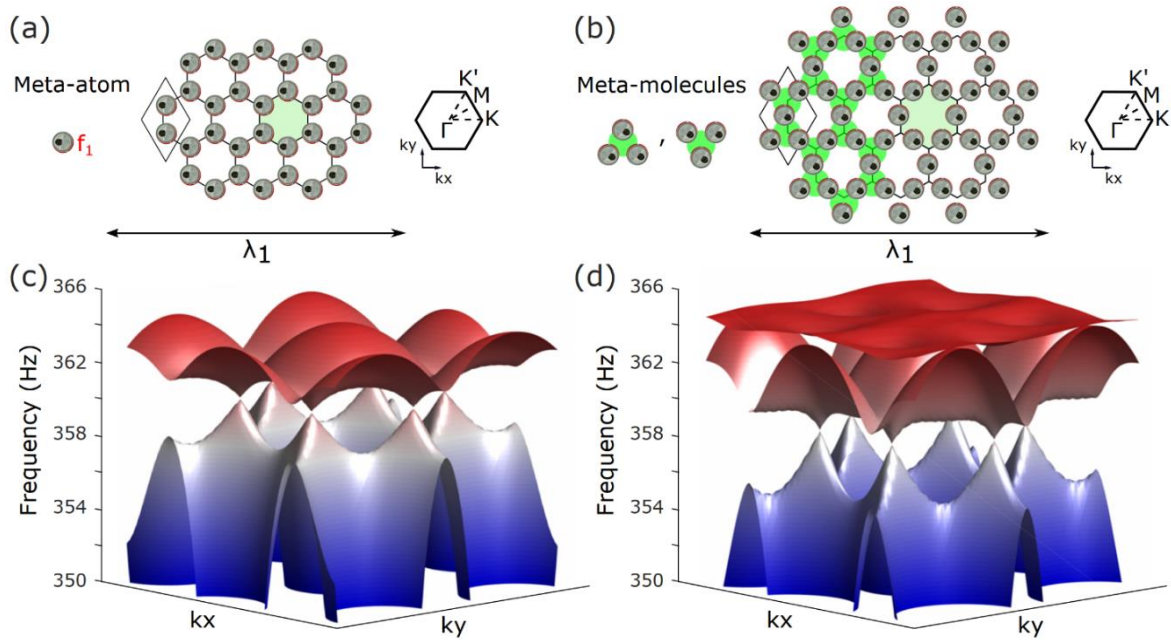

Fig. 1: (Colour on-line) (a) Subwavelength-scaled honeycomb lattice made of soda cans as meta-atoms and its first Brillouin zone (right). (b) Subwavelength-scaled Kagome lattice, which can be seen as a Honeycomb lattice of meta-molecules (green disks), and its first Brillouin zone (right). (c) Band diagram of the Honeycomb lattice. (d) Same as (c) for the Kagome lattice.

resonator whose resonant frequency $f_1$ is linked to the air volume it contains [19]. For the case of empty cans, $f_1$ is roughly 415 Hz, meaning a resonant wavelength around 80 cm which is far bigger than the size of a single can, making them relevant meta-atoms for our system. Building a subwavelength honeycomb lattice with these cans, as shown in Fig. 1(a), allows to obtain Dirac cones at the corners of the first Brillouin zone [39] (K and K' valleys) as we can see on the corresponding numerical dispersion relation displayed in Fig. 1(c) obtained with COMSOL Multiphysics. Interestingly, these peculiar degeneracies, which are the starting point before obtaining intriguing properties, also exist at the K points of the band diagram for the case of another hexagonal Bravais lattice, the Kagome lattice of cans (Fig. 1(b)), as depicted on Fig. 1(d). These two lattices, although different by nature because they do not contain the same number of meta-atoms per unit cell, are both hexagonal Bravais lattices and consequently they have some similarities, as for example the shape of the first Brillouin zone. It is therefore interesting to push the analogy between the two lattices, and to describe the Kagome lattice as a honeycomb lattice made of meta-molecules (made of a trimer of cans), as shown in green disks on Fig. 1(b), even if the latter are more complicated than the original meta-atoms. Each meta-molecule exhibits three resonances, a monopolar one and two dipolar ones, which are responsible for the propagating bands. This vision therefore permits to explain genuinely the presence of Dirac cones linking the two low-frequency bands in the Kagome lattice dispersion. Moreover, the existence of the third band is simply related to the two orthogonal symmetries of the trimer's dipolar modes: one of them acts as a deaf-mode and hybridizes poorly with the incoming propagating wave according to its direction, which generates this flat band. In the following we use this clear connection between the two lattices to create an acoustic analogue of the quantum valley-Hall effect with our metamaterial.

**Composition-structure connection to induce valley differentiation.** – In a previous work in the microwave domain [21], and recently with soda cans in acoustics [22], it has been proven that it is possible to break appropriately the Dirac cones of the honeycomb lattice by slightly detuning one of the two resonators of the unit-cell. In the case of the acoustic sample, it is realized by diminishing the Helmholtz cavity which is done by adding some water in one can in order to increase its resonance frequency $f_2 > f_1$. In that case we play with the composition of the unit cell, and the structure is left unchanged, as it is shown in Fig. 2(a). Now, exploiting the correspondence established in the previous paragraph, we can also detune one of the two meta-molecules of the Kagome unit-cell. However, this time we do not pour some water in the cans, one can being shared by the two meta-molecules but will play on the spatial pattern. Indeed, to open a band gap, we slightly modify the structure of the lattice as explained on Fig. 2(b). Notably, we shrink the trimer of one metamolecule (cyan disk) which directly expands the other trimer (purple disk). The resulting lattice is sometimes named breathing Kagome lattice and present higher-order topological features such as corner modes [40, 41, 42]. In our study, we show that this particular lattice can now directly be seen as a honeycomb lattice of two different meta-molecules: a small and a big trimer of soda cans. Playing with their spatial arrangement, we control the trimers' mode frequencies. More precisely, shrinking the trimers tends to get the corresponding monopolar and dipolar modes' frequencies slightly closer whereas expanding them results in the opposite behaviour.

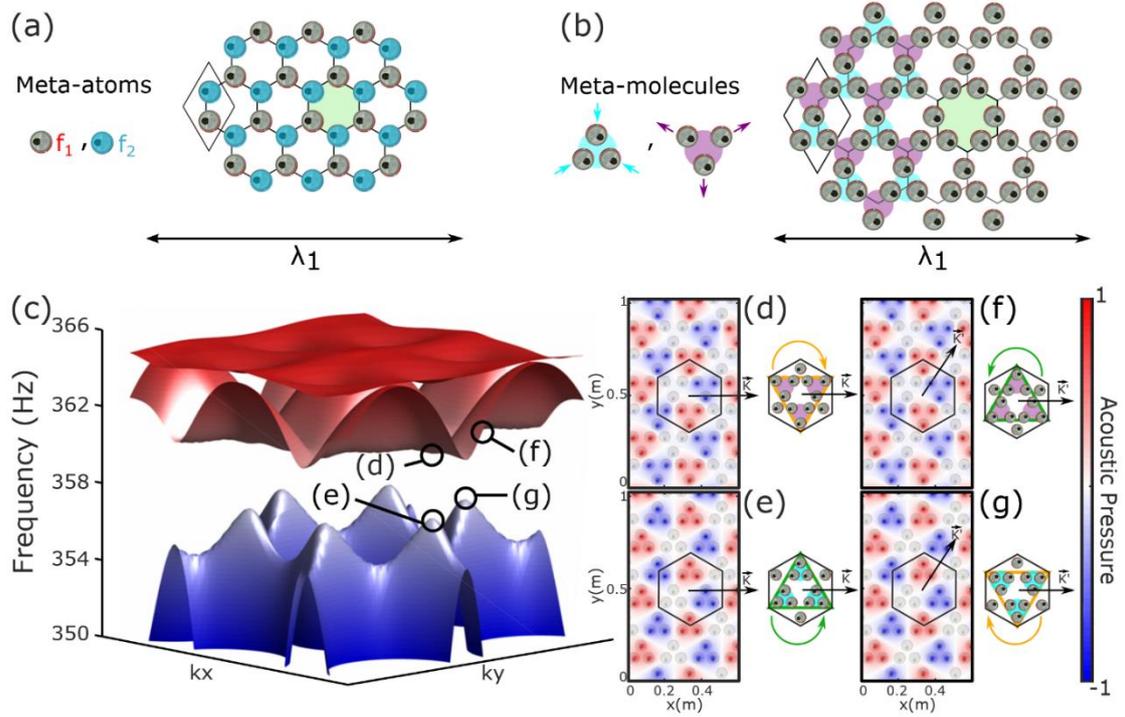

Fig. 2: (Colour on-line) (a) Bidisperse honeycomb lattice of soda cans with two different meta-atoms: empty and partially filled cans (blue). (b) Breathing Kagome lattice which can be seen as a honeycomb lattice of two different metamolecules: shrunk (cyan) and expanded (purple) trimers. (c) Dispersion relation of the breathing Kagome lattice of cans. (d) (resp. (e)) Acoustic field map of the crystalline mode at the valley K for the high frequency band (resp. low frequency band). Insets on the right picture the rotation of the field within the hexagon (dark line) (f) (resp. (g)) Same as (d) (resp. (e)) at the valley K'.

To demonstrate the coherence of our approach, we compute the numerical dispersion relation of the breathing Kagome lattice shown in Fig. 2(c) with COMSOL Multiphysics. The Dirac cones are replaced by a small bandgap. We do not worry about the wideness of the gap in the context of this article, but one has to remember that it is related to the changes made to the original Kagome lattice: the more the lattice is deformed the wider the gap. We now focus on the wave behaviour in the metamaterial at the K valley for the lower frequency band, whose corresponding field map is displayed on Fig. 2(e). In particular, we take a closer look at the pressure profile within a big hexagon (dark line). We clearly see that the pressure is distributed on a triangle of shrunk metamolecules (in green on the right of the frame). Moreover, as the wave propagates in the K direction this field distribution rotates on the green triangle with a left-handed polarization. For the high frequency band which is depicted on Fig. 2(d), the pressure profile is distributed on a triangle of expanded metamolecules (in yellow on the right of the frame), and also rotates but with a right-handed polarization. Then we look at the field behaviour at the other K' valley. For the lower band, the field is still distributed on the triangle of shrunk metamolecules, as shown in Fig. 2(g). However, if we compare to the previous case, the latter is now upside down with respect to the propagation direction (in yellow on the right of the frame) which results in a right-handed rotation. Persistently, the pressure field related to the high-frequency band has a left-handed rotation (Fig. 2(f)).

These simulated results are in complete agreement with the behaviour obtained in the case of a bi-disperse honeycomb lattice [21, 22]. In addition to undoubtedly prove the relevance of our approach, it shows that playing with the structure of our unit cell is equivalent to control its composition. We insist on the fact that these correspondences emerge thanks to the crystalline description of locally resonant metamaterials. This allows us to directly induce a bulk circular polarization (referred as valley degree of freedom in condensed matter physics) within the lattice of soda cans simply by modifying the structure of the medium at a subwavelength level.

**Subwavelength Acoustic valley-Hall effect.** – We now want to obtain an acoustic analogue of the quantum valley-Hall effect. The usual procedure to follow is to make an interface between two media on which opposite valleys are superposed. This can be done simply by applying a mirror symmetry with respect to the KK' direction within our breathing Kagome lattice of soda cans. It results in two different interfaces displayed on Fig. 3(a) and (e). First, we focus on the red interface (Fig. 3(a)). Its dispersion relation is shown in Fig. 3(b) and effectively proves the presence of a new band within the band gap of the medium. Looking at the pressure field map corresponding to the mode propagating towards the positive $x$, ie. with a positive group velocity (red dot on Fig. 3(b)), we clearly see that the wave is guided along the interface and present an antisymmetric profile with respect to the interface. Following the study made in microwaves [21], we can define a hexagon (dark line) which is now composed of two hybrid triangles (green and yellow) of shrunk and expanded meta-molecules as shown in Fig. 3(d). The field pressure is now distributed on the two triangles (green and yellow) and it is possible to define two counter-propagating circular polarizations as in the previous bulk study. Their sense of rotation is directly linked to the direction of propagation. In the case of the blue interface, a guided mode also lays in the bandgap (Fig. 3(f)). The field map of the mode going towards the positive x (blue dot on Fig. 3(f)) has now a symmetric profile with respect to the interface. It can also be described in terms of two opposite circular polarizations on different hybrid triangles at the interface, as presented in Fig. 3(g). Interestingly, these senses of rotation, which are related to the direction of propagation, are opposite compare to the red interface.

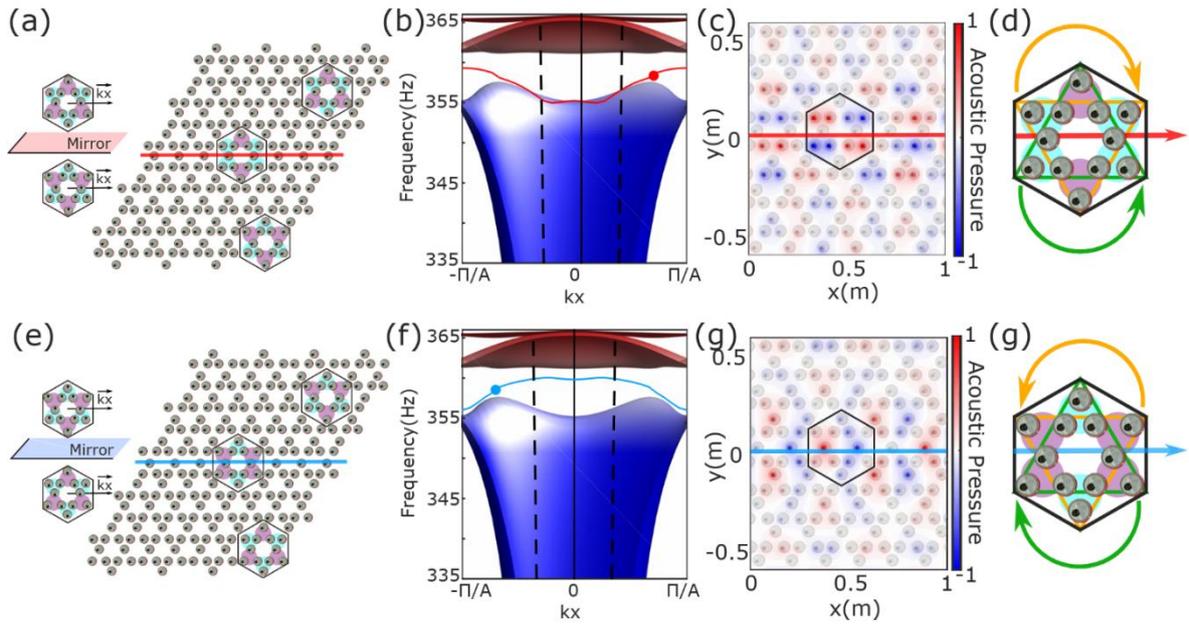

Fig. 3: (Colour on-line) (a) Red interface obtained by applying a mirror symmetry with respect to the KK' axis to the metamaterial. (b) Dispersion relation of (a). (c) Acoustic field map related of the guided mode corresponding to the red dot in (b). (d) Picture explaining the combined rotation scheme of the acoustic field on the resonators of the interface (dark lined hexagon of (c)). (e) Same as (a) but for the blue interface. (f) Dispersion relation of (e). (g) (resp. (g)) same as (c) (resp.(d)) for the blue interface.

Hence, these numerical simulations truly evidence that we managed to obtain an acoustic analogue of the quantum valley-

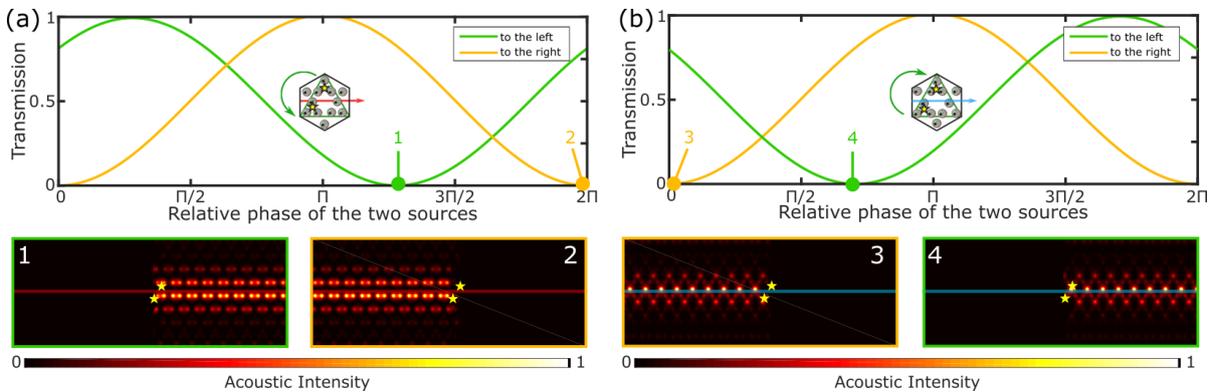

Fig. 4: (Colour on-line) (a) Transmission towards the left (green) and towards the right (yellow) of the two sources (stars in inset) according to the phase shift between them, in the case of the red interface. It is possible to excite unidirectionally the guided wave as shown on the intensity maps 1 and 2 (below). (b) Same as (a) for the blue interface. It is also possible to excite unidirectionally the guided wave as shown on the intensity maps 3 and 4 (below).

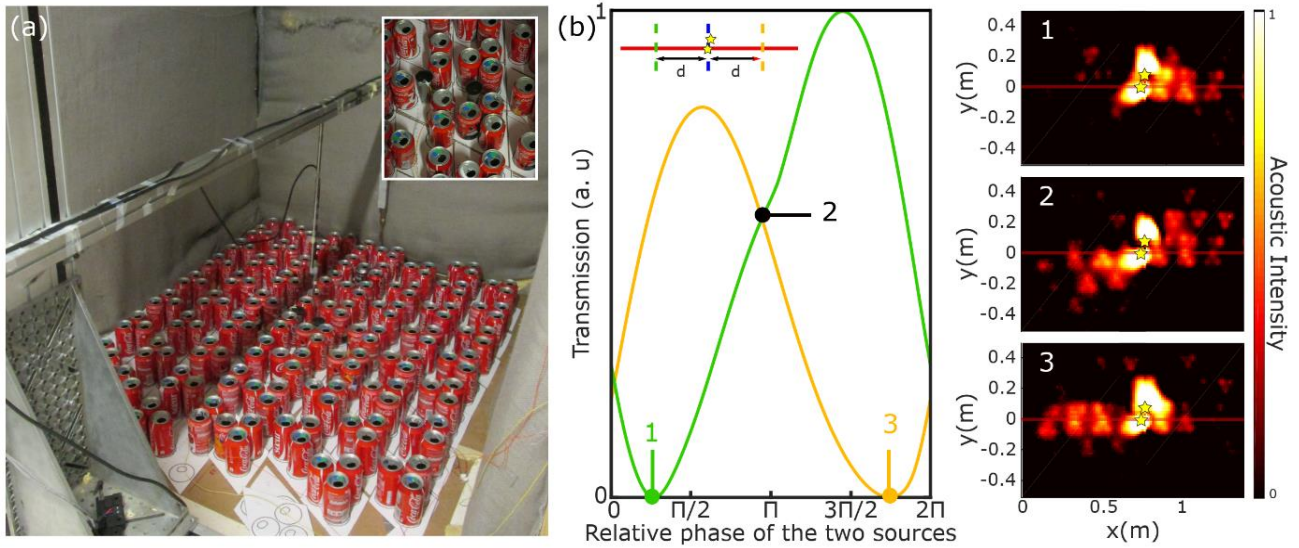

Fig. 5: (Colour on-line) (a) Experimental setup for the red interface. Zoom on the sources (inset). (b) Transmission results taken at a distance d on the left (green) and on the right (yellow) of the sources (yellow stars in inset), according to the phase shift between them. On the left, three experimental acoustic intensity maps show different cases: unidirectional to the right (1), omnidirectional (2), unidirectional to the left (3).

Hall effect with our structured lattice of soda cans. Particularly, we created two different subwavelength waveguides carrying an additional degree of freedom which we fully characterized as the combination of two counter-rotating patterns locally displayed on the interface metamolecules.

**Numerical study of the directional excitation of the guided wave.** – In the following, we exploit this intriguing property of the guided waves in order to make them propagate preferentially in one direction. As we know precisely where the modes are rotating spatially, we can therefore induce a rotational excitation simply by placing two sources at appropriate positions and by applying a phase shift between them. For instance, relevant locations are the corners of one of the two hybrid triangles described above in Fig. 3(g) and (d). From now, we focus on the green triangle and only represent it for matters of clarity. Nevertheless, one must keep in mind that the actual mode presents a combination of the two rotations linked to its propagation direction.

We first verify this idea numerically. The transmission dependence towards the phase shift between the two sources is presented on Fig. 4(a) and (b) for the case of the red and blue interfaces respectively. The sources' positions are indicated with yellow stars in the insets. For the red interface, we can see that the phase shift between the sources is responsible for a directional propagation along the domain wall. Moreover, it is possible to obtain purely unidirectional excitations for appropriate phase shifts between the two sources. Indeed, the field maps corresponding on these specific cases are depicted below as cases ranging from 1 to 4. This clearly proves that the sense of the field rotation on the green triangle is not only tightly linked to the direction of propagation of the wave but also that it is opposite between the two interfaces (case 3 and 4 have opposite orders compared to cases 1 and 2).

**Experimental demonstration of directional excitation of sound on a sub-wavelength scale.** – We now go further and implement this procedure experimentally, in the case of the red interface. The experimental setup is displayed on Fig. 5(a). We place the medium in an anechoic chamber and excite it with two loudspeakers placed at the previously described positions (see inset). The emitted signal is a chirp containing a broad frequency range ($\Delta f = 800$ Hz) centred on 415 Hz. A microphone, mounted on a two-dimensional translational stage, measures the acoustic field above the metamaterial, in its close vicinity. This way, we measure acoustic field maps associated to each source for each frequency. In a post-processing step, after having identified the frequencies of the guided modes along the interface, we apply a phase shift between the maps related to each loudspeaker and measure the transmission at a distance d = 40 cm on each side of the sources. The results are displayed on Fig. 5(b) (the procedure is in inset). They show an agreement with the previous numerical study. Notably, it is possible to excite unidirectionally the wave propagating along the domain wall as it is undoubtedly evidenced by the intensity maps related to the cases 1 and 3. Moreover, we insist on the fact that the resonant frequency of a soda can is $f_1 = 415$ Hz which corresponds to a wavelength ($\lambda_1 = 80$ cm) far bigger than the spatial scale of the guided wave. Hence, simply by controlling the phase of two loudspeakers we clearly managed to control and measure unidirectional excitation of sound guided at a subwavelength scale.

**Conclusion.** – To conclude, in this work we exploit the crystalline description of locally resonant metamaterials to obtain experimentally an acoustic analogue of the quantum valley-Hall effect with a lattice of soda cans. This microscopic approach of metamaterials allows to establish strong parallels between different lattices to induce field rotations within the medium simply by modifying its subwavelength structure. We then use it in order to guide waves along specific domain walls whose direction of propagation is tightly linked to the rotation of sound at the level of the interface. We exploit this singular property to send sonic waves preferentially in one direction by tailoring the excitation in a very simple manner. This numerical work and experimental verification clearly show that crystalline metamaterials have a great potential as a table-top platform to investigate tantalizing solid-state physics phenomena.

\*\*\*

S.Y. acknowledges funding from French Direction Générale de l'Armement. This work is supported by LABEX WIFI (Laboratory of Excellence within the French Program "Investments for the Future" ) under references ANR-10-LABX-24 and ANR-10-IDEX-0001-02 PSL\* and by Agence Nationale de la Recherche under reference ANR-16-CE31-0015.